\documentclass[manuscript,nonacm]{acmart}

\usepackage{booktabs}
\usepackage{tabularx}
\usepackage{array}
\usepackage{xcolor}
\usepackage{graphicx}
\usepackage{listings}
\usepackage{microtype}
\usepackage{tikz}
\usetikzlibrary{arrows.meta,calc,shapes.geometric}

\definecolor{paiUser}{HTML}{2F5D8A}
\definecolor{paiUserLight}{HTML}{E9F1F8}
\definecolor{paiAgent}{HTML}{A34A3A}
\definecolor{paiAgentLight}{HTML}{F8ECE8}
\definecolor{paiDecision}{HTML}{397261}
\definecolor{paiDecisionLight}{HTML}{E7F2EE}
\definecolor{paiBorder}{HTML}{C9CDD2}
\definecolor{paiNeutral}{HTML}{68717A}
\definecolor{paiPanel}{HTML}{FAFAF8}
\definecolor{paiRevision}{HTML}{B3261E}

\newcommand{\paiFigureFont}{\sffamily\fontsize{8.25}{9.9}\selectfont}
\newcommand{\paiPanelTitleFont}{\sffamily\bfseries\fontsize{9}{10.8}\selectfont}
\newcommand{\paiSecondaryFont}{\sffamily\fontsize{7.5}{9}\selectfont}
\newcommand{\paiAvatarFont}{\sffamily\bfseries\fontsize{7}{8.4}\selectfont}

\tikzset{
  pai-figure/.style={
    x=1cm,
    y=1cm,
    font=\paiFigureFont
  },
  pai-panel/.style={
    rounded corners=5pt,
    fill=paiPanel,
    draw=paiBorder,
    line width=0.55pt
  },
  pai-panel-title/.style={
    font=\paiPanelTitleFont
  },
  pai-chat-surface/.style={
    rounded corners=4pt,
    fill=white,
    draw=paiBorder,
    line width=0.55pt
  },
  pai-user-message/.style={
    rounded corners=3pt,
    draw=paiUser,
    fill=paiUserLight,
    line width=0.7pt,
    minimum height=0.46cm,
    inner xsep=5pt,
    inner ysep=3pt,
    align=center
  },
  pai-agent-message/.style={
    rounded corners=3pt,
    draw=paiAgent,
    fill=paiAgentLight,
    line width=0.7pt,
    minimum height=0.46cm,
    inner xsep=5pt,
    inner ysep=3pt,
    align=center
  },
  pai-user-avatar/.style={
    circle,
    draw=paiUser,
    fill=paiUser,
    text=white,
    line width=0.6pt,
    minimum size=0.46cm,
    inner sep=0pt,
    font=\paiAvatarFont
  },
  pai-agent-avatar/.style={
    circle,
    draw=paiAgent,
    fill=paiAgent,
    text=white,
    line width=0.6pt,
    minimum size=0.46cm,
    inner sep=0pt,
    font=\paiAvatarFont
  },
  pai-secondary/.style={
    text=paiNeutral,
    font=\paiSecondaryFont
  },
  pai-divider/.style={
    densely dotted,
    draw=paiBorder,
    line width=0.55pt
  },
  pai-silence-span/.style={
    densely dotted,
    draw=paiNeutral,
    line width=0.65pt
  },
  pai-user-event/.style={
    rounded corners=2pt,
    draw=paiUser,
    fill=paiUserLight,
    line width=0.7pt,
    minimum width=0.78cm,
    minimum height=0.45cm,
    inner sep=2pt
  },
  pai-agent-event/.style={
    rounded corners=2pt,
    draw=paiAgent,
    fill=paiAgentLight,
    line width=0.7pt,
    minimum width=0.78cm,
    minimum height=0.45cm,
    inner sep=2pt
  },
  pai-pending-event/.style={
    rounded corners=2pt,
    draw=paiAgent,
    fill=white,
    dashed,
    line width=0.8pt,
    minimum width=0.9cm,
    minimum height=0.48cm,
    inner sep=2pt
  },
  pai-plan-group/.style={
    rounded corners=3pt,
    draw=paiDecision,
    fill=paiDecisionLight,
    densely dashed,
    line width=0.6pt
  },
  pai-decision/.style={
    diamond,
    aspect=1.45,
    draw=paiDecision,
    fill=paiDecisionLight,
    line width=0.8pt,
    minimum width=0.65cm,
    minimum height=0.58cm,
    inner sep=1pt
  },
  pai-event-line/.style={
    draw=paiBorder,
    line width=0.45pt
  },
  pai-causal/.style={
    -{Latex[length=1.7mm]},
    draw=paiNeutral,
    line width=0.55pt
  },
  pai-revision/.style={
    -{Latex[length=1.7mm]},
    draw=paiRevision,
    line width=0.7pt
  },
  pai-axis/.style={
    -{Latex[length=1.8mm]},
    draw=paiNeutral,
    line width=0.55pt
  },
  pai-cancel-mark/.style={
    draw=paiRevision,
    fill=white,
    text=paiRevision,
    circle,
    line width=0.7pt,
    minimum size=0.34cm,
    inner sep=0pt,
    font=\paiAvatarFont
  }
}

\setcopyright{none}
\renewcommand\footnotetextcopyrightpermission[1]{}
\hypersetup{pdfauthor={Zijie Yang}}

\definecolor{codegreen}{HTML}{164F45}
\definecolor{codecoral}{HTML}{A93428}
\definecolor{codegray}{HTML}{F2EFE8}
\lstdefinelanguage{json}{
  basicstyle=\ttfamily\footnotesize,
  backgroundcolor=\color{codegray},
  breaklines=true,
  columns=fullflexible,
  keepspaces=true,
  showstringspaces=false,
  string=[s]{"}{"},
  stringstyle=\color{codegreen},
  literate=
   *{0}{{{\color{codecoral}0}}}{1}
    {1}{{{\color{codecoral}1}}}{1}
    {2}{{{\color{codecoral}2}}}{1}
    {3}{{{\color{codecoral}3}}}{1}
    {4}{{{\color{codecoral}4}}}{1}
    {5}{{{\color{codecoral}5}}}{1}
    {6}{{{\color{codecoral}6}}}{1}
    {7}{{{\color{codecoral}7}}}{1}
    {8}{{{\color{codecoral}8}}}{1}
    {9}{{{\color{codecoral}9}}}{1}
}
\newcolumntype{Y}{>{\raggedright\arraybackslash}X}

\title{Proactive AI: From Turns to Replannable Dialogue Timelines}

\author{Zijie Yang}
\affiliation{%
  \institution{Airalogy}
  \city{Hangzhou}
  \country{China}
}
\email{yangzijie@airalogy.com}

\begin{abstract}
Conventional large-language-model chat interfaces typically follow user-initiated turns:
each request elicits a response, after which the system waits for further input. Human
asynchronous communication instead unfolds over time through message bursts,
delays, silence, resumed topics, and self-initiated contact. Proactive dialogue
therefore requires not only deciding when to speak, but also managing pending
conversational actions and allowing them to be revised as context evolves.
We introduce Proactive AI, a framework for proactive dialogue built around
replannable temporal message queues. The framework treats interaction as a
continuous event timeline and unsent conversational actions as revisable state.
A user event, dialogue-pacemaker event, or scheduled
decision may yield silence, one or more immediate messages, or future
conversational actions. The system may also observe a user message without an
immediate visible response and defer the response decision. Before delivery,
every planned message must be reconsidered against the current context; reaching
a scheduled time does not itself authorize delivery.
We provide a formal semantics that characterizes when future conversational
actions may be revised, withdrawn, or executed as context evolves. The framework
extends AI participation beyond responses to current requests, enabling
self-initiated exchanges without new requests, sustained follow-up over time,
and revisions of subsequent actions as context evolves. It provides an
executable basis for long-term human--AI interaction in tutoring, scientific
collaboration, and everyday companionship.
\end{abstract}



\begin{document}
\maketitle

\section{Introduction}

\begin{figure*}[t]
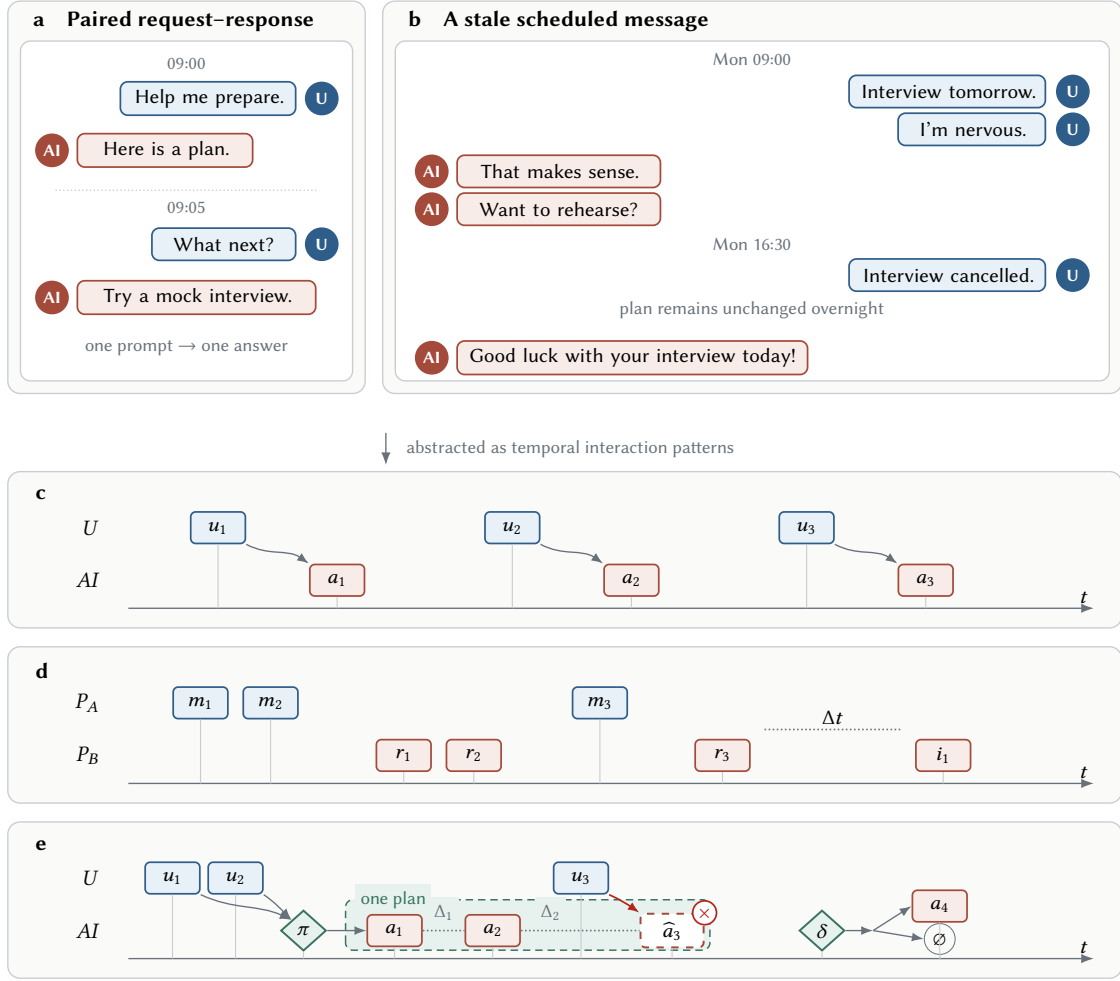

  \centering
  \resizebox{0.98\textwidth}{!}{\begin{minipage}{15.9cm}
  \centering
  \input{figures/conversation-examples.tex}

  \vspace{0.12cm}
  \begin{tikzpicture}[pai-figure]
    \draw[pai-axis]
      (0,0.24) -- (0,-0.22);
    \node[pai-secondary, anchor=west] at (0.18,0.01)
      {abstracted as temporal interaction patterns};
  \end{tikzpicture}
  \vspace{0.04cm}

  \input{figures/interaction-timelines.tex}
\end{minipage}}
  \caption{From concrete conversations to temporal interaction patterns.
  \textbf{a},~A conventional chat alternates one user request and one assistant answer.
  \textbf{b},~A failure case: after the user reports that the interview was cancelled,
  the AI still sends an encouragement planned the previous day. In the timelines
  below, $t$ denotes wall-clock time.
  \textbf{c},~Paired chat proceeds through alternating user inputs and AI answers.
  $u_i$ denotes the $i$th user message and $a_i$ its paired AI answer.
  \textbf{d},~Human asynchronous communication allows both participants to send
  multiple consecutive messages and either participant to resume contact after
  silence. $P_A$ and $P_B$ denote the two participants; $m_i$ and $r_i$ denote
  messages from $P_A$ and replies from $P_B$, respectively. After an interval
  $\Delta t$ without communication, $P_B$ initiates $i_1$ without waiting for a
  new message from $P_A$.
  \textbf{e},~The proposed model organizes multiple future messages into a temporal
  plan that can be revised as context changes, while allowing the system to
  reconsider whether to speak without new user input. Independently sent user
  messages $u_1,u_2$ enter one decision snapshot, from which planner $\pi$ schedules
  three messages: $a_1$ is delivered immediately, $a_2$ after interval $\Delta_1$,
  and $\widehat{a}_3$ has an earliest delivery time a further interval $\Delta_2$
  later but is cancelled by a preceding user update $u_3$. The later
  reconsideration $\delta$ may remain silent or create a new message plan;
  $a_4$ illustrates only the single-message case, and the new plan may instead
  contain multiple immediate or delayed messages.}
  \Description{A five-panel, two-level figure. Panels a and b are chat
  transcripts; panels c--e are horizontal timelines running from left to
  right. Blue rounded rectangles mark user or participant-A messages, and coral
  rectangles mark AI or participant-B messages. Panel c links each user message
  to one AI answer. Panel b shows an encouragement planned before the interview
  was cancelled being sent after it has become stale. Panel d shows bursts from
  both humans followed by a gap and a participant-B-initiated message. Panel e
  shows two independently sent user messages entering one decision snapshot. A green dashed group
  contains one immediate message, one delayed message, and one pending message,
  with $\Delta_1$ and $\Delta_2$ marking their intervals. A later user update
  cancels the pending message. A second green diamond then branches to either a
  representative single new AI message or an empty-set symbol for silence.}
  \label{fig:timeline}
\end{figure*}

The dominant interface for a large language model (LLM) is deceptively simple:
the user submits input, the model produces one visible assistant turn, and the
system then waits for another user action. This contract is useful for question
answering and command execution, yet it is a poor description of asynchronous
social messaging. People often divide a thought across several messages, respond
to a group of messages together, send multiple replies in succession, read a
message without responding immediately, return to an earlier topic, or initiate
a later exchange (Figure~\ref{fig:timeline}; Supplementary Figure~\ref{fig:dialogue-patterns}).

Among these phenomena, initiative exposes a fundamental limitation of turn-based
chat: once the user stops providing input, the system loses its trigger for
participating again. Even after both people end an exchange with ``goodbye,''
either person may reopen the conversation later; a system driven purely by user
turns instead waits indefinitely. A persistent proactive system therefore first
needs opportunities to deliberate without new user input. Yet this alone does
not guarantee that the conversation continues: if the system chooses silence at
every such opportunity, the interaction can still stall indefinitely.

Prior dialogue research has relaxed parts of this turn-based assumption. Proactive
dialogue systems may introduce content or steer toward a goal
\cite{li2016stalemate,wu2019duconv,deng2023survey}. Turn-taking systems decide
whether to speak or wait \cite{lan2019when,lin2020wait}. Response-timing models
predict natural offsets in spoken dialogue \cite{roddy2020timings}. Time-aware
models use gaps between sessions to choose better topics \cite{zhang2023gap}, and
persistent agents remember, plan, and initiate behavior
\cite{park2023generative,liu2025inner}. Recent work also supports burst-style
exchange in which each party may emit several consecutive timestamped messages
\cite{wu2025xturing}.
Together, these strands provide essential ingredients for asynchronous AI
dialogue, but treat initiative, silence, elapsed time, and message bursts as
separate capabilities rather than jointly revisable state on a shared timeline.

The interview scenario in Figure~\ref{fig:timeline}b shows how a message planned
in advance can become stale. When the user tells the AI that the interview is the
next day, the AI plans to send ``Good luck with your interview today!'' the next
morning. The user subsequently reports that the interview has been cancelled. If
the system does not re-examine the plan in light of this update, it still sends
the encouragement the next morning, producing proactive contact that conflicts
with the latest context. A pending message therefore cannot be treated as an
immutable scheduled task. On receiving new user input, the system should
re-evaluate the assumptions on which the plan depends and cancel, revise, or
reschedule it---or remain silent after reconsideration.

This raises a central question: \emph{how should a dialogue system manage
messages not yet sent and decisions scheduled for the future, and revise them
as circumstances change?} A fixed timer may release previously generated content
even after it has become obsolete. A heartbeat mechanism can wait until the
system wakes before deciding whether to respond, but it typically schedules only
the next decision. A model can instead generate a batch of timestamped
messages and replace the batch when replanning; this still makes it difficult to
retain, revise, postpone, or cancel individual plans independently. Asynchronous
dialogue therefore requires not only deciding what to say now, but also managing
what may happen later as subsequent events change those decisions.

To address this problem, we introduce Proactive AI, a framework for proactive
LLM conversations. Its core mechanism is the \emph{replannable temporal message
queue}, which treats conversation as a continuous event timeline and keeps
pending conversational actions as persistent,
revisable conversational state. User messages, dialogue-pacemaker events, and due
actions can all trigger a new plan. The planner may remain silent, send one or more
messages now, schedule future actions, or revise an earlier plan. Every future
action is reconsidered against the latest context before execution, so reaching a
scheduled time does not itself authorize delivery.

This temporal structure supports long-term interactions in which circumstances
evolve and new information may invalidate earlier plans. For example, an AI tutor
may revisit an unresolved learning difficulty after an interval, an AI scientific
assistant may follow up after an experiment or new result, and an AI companion
may draw on earlier exchanges to initiate a supportive check-in at an appropriate
time. These systems may also revise contact plans or remain silent in light of
the latest context.
Proactivity thereby becomes not an isolated trigger but an ongoing conversational
process that can be revised as circumstances change and remains subject to user
control.

This paper makes the following three contributions:

\begin{enumerate}
  \item We introduce a conversation model organized around a continuous event
  timeline rather than user turns, representing silence, multi-message bursts,
  delayed replies, and proactive initiation within a single framework. The model
  also supports a configurable bound on continued silence, allowing temporary
  silence while preventing indefinite waiting for user input due to repeated
  silence decisions.
  \item We formalize a replannable temporal message queue that distinguishes
  prepared $\mathsf{Send}$ actions from $\mathsf{Reconsider}$ actions that reserve
  a future decision point without fixing its wording. Multiple future intentions
  can therefore be retained, postponed, replaced, or cancelled independently.
  \item We require future actions to be reconciled with the latest context before
  execution and confirmed, revised, or cancelled within the same state update,
  preventing stale messages from being sent merely because their scheduled time
  has arrived.
\end{enumerate}

\section{Motivating Scenario and Design Requirements}

Consider a user who sends, ``I have an important interview tomorrow'' and ``I am
a little nervous'' as two messages. In one planning step, the system may schedule
an acknowledgment, an easy question after a short conversational interval, and
a concrete preparation suggestion 20 seconds later. Two hours later, the assistant may want
to check whether preparation or rest would now be more helpful. If the user
instead says ``the interview was cancelled'' before the 20-second message, the
prepared advice should disappear.

The system may also observe both messages without speaking immediately, leaving
whether and what to say to a later decision. We call this \emph{deferred response},
distinguishing it from network latency or delayed display of an already written
answer. For
example, after observing the two interview messages at 9:00, the system may say
nothing immediately and schedule a new decision for 9:10. If the context remains
unchanged, it might then reply, ``Feeling nervous is normal. Which part worries
you most?'' If the user adds ``the interview was cancelled'' at 9:05, the 9:10
decision must instead use the updated context; reassurance or preparation advice
that might previously have been appropriate should not appear unchanged.

Two superficially similar forms of waiting must be distinguished. First, the
planner may already have observed an earlier message and deliberately defer its
response by reserving a later reconsideration. Second, the user may send faster
than the model can infer, so new input arrives before an earlier candidate plan
has completed. The latter does not constitute a decision to remain silent; the
context underlying the older decision has changed before commit. We represent
the first case with $\mathsf{Reconsider}$ and handle the second through input
coalescing and decision-snapshot freshness checks.

Whether the system had scheduled a prepared message or only a later
reconsideration, the cancellation should trigger replanning under the updated
context. For example, the assistant might respond, ``I understand. Then you do
not need to prepare for the interview tonight. Would you like to talk about what
happened, or would you rather rest?'' The need for any later follow-up should
likewise be reconsidered. The point is to revise plans as circumstances change,
rather than prescribe a particular response.

Figure~\ref{fig:timeline}e shows how $a_1$, $a_2$, and $\widehat{a}_3$ are
created in one plan with distinct delivery times, and how a later user event
cancels an unsent action before execution. The required mechanism is not simply
delayed delivery: user events must interrupt and revise existing plans. For a
longer-horizon follow-up, the system should schedule future reconsideration and
decide at that time, using the then-current context, whether to generate a
message rather than fixing its wording in advance.

Another boundary case arises after an exchange has ended naturally. Suppose both
parties say ``goodbye'' and the user sends nothing further. A conventional chat
waits indefinitely. To let the system reconsider whether to open a new exchange
without new user input, we introduce a dialogue pacemaker. It periodically
wakes the planner and triggers a new decision, grounded in the then-current
context, about whether to initiate contact. The result may be one or more messages
or continued silence. The pacemaker therefore guarantees another opportunity to
decide, not that the system speaks. If the planner chooses silence at every
wakeup, the dialogue may still stall indefinitely. A dialogue pacemaker alone is
therefore insufficient: the system also needs a configurable bound on continuous
silence. Once that bound is reached, the next valid decision must initiate contact
from the then-current context rather than release a prewritten opener.

These scenarios yield five interaction requirements (R1--R5). System
implementation and deployment also require two execution and control conditions
(R6--R7).

\paragraph{R1: Planning need not be user-triggered, and message counts may vary.}
Every user send first becomes an independent arrival event and requests renewed
deliberation; nearby arrivals may be jointly observed in one decision snapshot
without requiring one model call per message. User messages, pacemaker pulses,
and due queue actions may all trigger planning. A planning step may observe zero,
one, or many new user messages and produce zero, one, or many immediate messages
and future actions. Silence, message bursts, and a deferred first visible
response are ordinary outcomes.

\paragraph{R2: Wall-clock time as system state.}
Future actions may span seconds to days. The system must explicitly represent
current time and each action's earliest eligibility time rather than relying on
turn order.

\paragraph{R3: Revisable pending actions.}
Unsent messages and scheduled reconsiderations are conditional intentions. Later
events may retain, postpone, replace, or cancel them independently.

\paragraph{R4: Decision-time generation for long-horizon follow-up.}
When a follow-up's premises may change over a longer interval, the system should
schedule future reconsideration rather than pre-generate the text to be sent.

\paragraph{R5: Silence is bounded in proactive mode.}
When bounded silence is enabled, the system must support a configurable deadline
for continuous silence. The next valid decision after that deadline, when external
rules permit contact, must
produce at least one visible message; the system cannot cancel or indefinitely
postpone that contact merely by choosing silence again.

\paragraph{R6: Separation of generative judgment and deterministic control.}
Models may judge content, relevance, and tone, but a deterministic runtime must
govern clocks, event order, state updates, and delivery authorization.

\paragraph{R7: Inspectable actions under external policy.}
Pending actions, action rationales, wakeups, and silence decisions must be
inspectable. Autonomous contact must respect user authorization, quiet hours,
and rate limits as determined by external policy. Users must be able to revoke
pending contact arrangements.

\section{Related Work}

\subsection{Mixed Initiative and Goal-Directed Dialogue}

Mixed-initiative interaction combines system intervention with direct user
control. Horvitz argues that decisions about whether and when to act should
account for uncertainty about user goals, the costs and benefits of intervention,
and the user's attention \cite{horvitz1999mixed}. This perspective provides a
foundation for the timing and control of proactive contact; we further examine
how multiple future conversational actions formed in one decision can persist
and be revised when later events arrive.

StalemateBreaker detects conversational stalemate and introduces externally
sourced content \cite{li2016stalemate}. Its motivation overlaps with ours: both
reject the assumption that the user must always lead. Its mechanism, however,
reacts to semantic signs of stalemate within an ongoing session and selects new
content; it does not specify wall-clock wakeups or address when the system should
reinitiate contact after repeatedly choosing silence. DuConv gives one participant
an explicit knowledge-graph path and asks it to lead discussion toward a goal
\cite{wu2019duconv}.
The broader proactive-dialogue literature studies topic guidance, clarification,
persuasion, recommendation, and other system-side goals \cite{deng2023survey}.
These systems primarily alter the content or direction of an active dialogue
policy. We instead study how linguistic actions that may occur later are
represented and revised, including after user input has ceased.

Proactive Agent, proposed by Lu et al., uses user activities and environmental
events to identify potential needs and offer assistance without an explicit
request \cite{lu2025proactive}. Its focus is on decisions about proactive
intervention; we examine how future conversational actions persist across
successive decisions and are revised or cancelled as new information arrives.

\subsection{Waiting, Bursts, and Response Timing}

When to Talk augments a multi-turn generator with a decision to continue speaking
or stay silent \cite{lan2019when}. Wait, I'm Still Talking! begins from the
instant-messaging observation that users split one thought across several short
messages; its Imagine-Then-Arbitrate model predicts whether to wait for more
input or respond now \cite{lin2020wait}. These works are direct precedents for
non-mandatory response and input batching. Their silence decisions are local to
the current interaction point: they neither maintain a long-horizon future queue
nor address whether contact will resume after repeated silence decisions.

X-TURING implements burst dialogue with batched user input, model-generated
consecutive replies, timestamp-governed delivery, and removal of still-unsent
messages when a new response batch is produced \cite{wu2025xturing}. These
mechanisms support a long-term Turing-test framework for evaluating human
likeness. We ask whether multiple future actions can persist across separate
decisions and be retained, edited, postponed, or cancelled individually; reconsideration
can also be scheduled without prewriting speech. The distinction is therefore
not multi-message output itself, but keeping future messages and decisions
available for later revision.

Roddy and Harte learn context-sensitive spoken response offsets and find that
listeners' naturalness judgments depend on timing \cite{roddy2020timings}. That
work assumes a dialogue manager has already decided to take a turn and predicts
fine-grained spoken onset. We address asynchronous text over longer horizons and
jointly represent whether, what, and how many messages may occur.

\subsection{Persistent Intentions and Time-Aware Agents}

The persistence and revision of intentions are central concerns in classical
agent research. Bratman et al. treat plans as partial commitments that constrain
subsequent reasoning, avoiding the need to decide everything afresh at every
moment \cite{bratman1988plans}. Cohen and Levesque formalize commitment in
intention and the conditions under which changing beliefs permit goals to be
dropped \cite{cohen1990intention}. These accounts explain why future action
requires both persistence and revisability. We address this problem in
asynchronous conversation by representing future messages and reconsideration
opportunities, and specifying their treatment when new events arrive and when
execution becomes due.

GapChat supplies elapsed gaps and expected event progress so a model can select
appropriate topics across sessions \cite{zhang2023gap}. Time affects generation,
but does not independently wake the model or mutate future speech. Generative
Agents combine memory, reflection, and daily planning in a simulated world and
can initiate conversations \cite{park2023generative}. Their planning object is
general behavior rather than a human--AI messaging protocol.

Inner Thoughts generates covert candidate thoughts, evaluates motivation to
express them, and seeks an
appropriate moment to participate in multi-party conversation
\cite{liu2025inner}. The public framing emphasizes a continuous internal process
and next-speaker selection. We further ask how multiple future utterances can
persist over wall-clock time and be updated together as circumstances change,
including whether contact should resume after repeated decisions not to speak.

\subsection{Heartbeat-Based Proactive Agents}

The open-source ProactiveAgent library records dialogue history, dynamically
chooses a sleep interval, wakes, and decides whether to generate a response; new
user input interrupts sleep, and its scheduler can bound the interval between
responses \cite{mariga2024proactiveagent}. We place the silence bound and multiple
future conversational actions on the same replannable timeline, so later events
can revise, postpone, or cancel them before delivery under external policy.

Taken together, prior work provides foundations for system-initiated speech,
waiting, multi-message output, temporal context, periodic autonomous wakeups,
and response-interval control. We place these capabilities on a shared event
timeline, where future messages, reconsiderations, and silence bounds evolve
together with user input, elapsed time, and external policy.

\section{Temporal Conversation Model}

To provide a unified account of deferred responses, message bursts, revisable
future actions, and system initiation without new user input, we formalize the
temporal conversation model underlying Proactive AI. The model treats events in wall-clock time rather
than request--response turns as the basic unit of interaction, and represents
unsent messages and future reconsiderations as revisable dialogue state. User
input, future-action eligibility, and periodic wakeups can each trigger a state
transition.

This choice allows a dialogue to evolve without further user input and allows
unsent content to be revised against new context before delivery. In
Figure~\ref{fig:timeline}e, a pending interview suggestion becomes obsolete
after new user input, providing an intuitive instance of the latter change.

\subsection{Event Timeline and Dialogue State}

At wall-clock time $t$, let $S_t$ denote system state:

\begin{equation}
S_t=\langle H_t,Q_t,P_t,L_t\rangle.
\end{equation}

Here, $H_t$ is visible message history with timestamps and processing markers, $Q_t$ is persistent future-
action state, $P_t$ records the next dialogue-pacemaker wakeup, and $L_t$ is
proactive-contact policy. Angle brackets $\langle\,\rangle$ denote an ordered
tuple: they combine components with fixed positions and meanings rather than
forming an unordered set. The same notation is used below for event bundles,
planner outputs, and proactive-contact policy.

Every user send becomes an independently timestamped message $u_i$ and is
immediately appended to $H_t$. An arrival requests a new decision opportunity,
but does not itself determine when a model call occurs. When the runtime takes a
decision snapshot at time $t$, it orders all messages that have arrived but have
not yet been handled by a valid plan. Let $E_t$ denote the event bundle jointly
observed by the resulting transition:

\begin{equation}
E_t=\langle U_t,w_t,D_t\rangle,
\end{equation}

where $U_t$ is the ordered list of zero or more independently arrived user
messages not yet handled by a successfully committed plan when the snapshot was
taken. Messages are ordered by arrival time and stable identifier. Reading a
message does not complete its processing: its marker in $H_t$ is updated only
when the corresponding plan commits, including a valid plan that remains silent.
A failed plan or one rejected because its context changed leaves these markers
unchanged, so previously read messages may enter a later $U_t$. We use square brackets for
ordered lists and $[]$ for an empty list.
$w_t\in\{0,1\}$ indicates whether the dialogue pacemaker fires. For a future
action $q$ in $Q_t$, $\operatorname{status}(q)=\textit{pending}$ means that it
has not been delivered, consumed, or cancelled, and $\tau(q)$ is its earliest
eligibility time. The set of actions that are eligible but unresolved at time
$t$ is therefore

\begin{equation}
D_t=\{q\in Q_t\mid \operatorname{status}(q)=\textit{pending}
\land \tau(q)\leq t\},
\end{equation}

where $\mid$ means ``such that'' and $\land$ denotes logical conjunction: both
conditions must hold. A transition contains at least one new user message, a
pacemaker wakeup, or a due action. User messages that arrived before the decision
snapshot and due actions are observed by one plan, so the system cannot
mechanically release a due message before reading an update that changes its
context.

The relation among state, event bundle, and updated state can be summarized as

\begin{equation}
S_t \xrightarrow{\;E_t\;} S_{t^+}.
\end{equation}

Here, $t^+$ denotes the point immediately after the events at time $t$ have been
processed. The system begins in state $S_t$, processes event bundle $E_t$, and
enters updated state $S_{t^+}$. Determining how this transition occurs requires
semantics for the future actions in $Q_t$ and constraints on the plans proposed
by the planner.

\subsection{Semantic Types of Future Action}

Future actions should not all be stored in the same form. For an utterance that
will follow after a short interval, the system may already have formed specific
wording. For a follow-up hours or days later, however, generating the message in
advance makes it vulnerable to changes in context; retaining an opportunity to
decide later is more appropriate. We therefore define two types of future
action. A $\mathsf{Send}$ action stores candidate text that is expected to
remain useful over a short interval, but still requires confirmation against the
latest context before delivery. A $\mathsf{Reconsider}$ action stores no message
content. At a specified time, it restores the current context and lets the
system decide whether to send one or more messages or remain silent. It can
represent either a later proactive follow-up or a deferred first response to
user input. The former type preserves an expression awaiting confirmation,
whereas the latter preserves a decision opportunity before any expression has
been fixed.

By definition, a future action occurs after the event that creates it; content
produced during the same event is immediate output. For an action $q$ created at
time $t$, the planner therefore proposes a positive delay $d(q)>0$. The
deterministic runtime checks that delay against permitted bounds and computes the
earliest eligibility time

\begin{equation}
\tau(q)=t+d(q).
\end{equation}

For example, the prepared short-delay suggestion in
Figure~\ref{fig:timeline}e can be represented by $\mathsf{Send}$, whereas a
decision about whether to follow up hours later is better represented by
$\mathsf{Reconsider}$.

To distinguish revisable future actions from actions that have already ended,
each action also records a stable identifier, creation event, reason, and status.
Its lifecycle is

\begin{equation}
\begin{aligned}
\textit{pending} &\rightarrow \textit{delivered} &&\text{accepted }\mathsf{Send},\\
\textit{pending} &\rightarrow \textit{consumed} &&\text{accepted }\mathsf{Reconsider},\\
\textit{pending} &\rightarrow \textit{cancelled} &&\text{cancellation},\\
\textit{pending} &\rightarrow \textit{pending} &&\text{replacement or rescheduling}.
\end{aligned}
\end{equation}

Delivered, consumed, and cancelled actions no longer participate in later
planning. Replacement or rescheduling changes an action's content or time while
leaving it pending. These states prevent the same action from being delivered or
processed more than once.

\subsection{Temporal Plans and Visible Output}

State $S_t$ and event bundle $E_t$ describe what the system knows at time $t$,
but not what it should do next. It must still decide whether to speak
immediately, schedule future actions, or revise existing plans. We use $\pi$ to
denote the planner that makes this decision. Given current state $S_t$ and event
bundle $E_t$,
$\pi(S_t,E_t)$ denotes the candidate temporal plan it proposes:

\begin{equation}
\pi(S_t,E_t)=\langle I_t,N_t,\Delta_t,A_t\rangle,
\end{equation}

where $I_t$ is an ordered list of immediate messages, $N_t$ is a set of new
future actions, $\Delta_t$ is a set of operations that revise existing actions,
and $A_t$ is a set of due-action identifiers explicitly accepted by this plan.
Each component may be empty, and $I_t$ may contain one or more messages. A single
plan can therefore represent silence, a single reply, or a burst of replies.

To separate proposed linguistic actions from system-state updates, we define the
transition function $\mathcal{T}$. It takes current state $S_t$, event bundle
$E_t$, and planner output $\pi(S_t,E_t)$ as inputs, and produces a successor
state only when the candidate plan satisfies the model constraints:

\begin{equation}
S_{t^+}=\mathcal{T}\bigl(S_t,E_t,\pi(S_t,E_t)\bigr).
\end{equation}

This separation assigns candidate-action generation to $\pi$ and constrained
state evolution to $\mathcal{T}$, preventing incomplete or conflicting planner
output from directly changing system state.

Let $V_t$ denote the list of messages ultimately shown to the user by this state
transition. It contains the immediate messages in $I_t$ and the prepared messages
whose $\mathsf{Send}$ actions have reached their earliest delivery time and are
confirmed by the current plan. Recording $A_t$ explicitly separates ``the time
has arrived'' from ``delivery is confirmed against the latest context.'' More
precisely, the text stored by a $\mathsf{Send}$ action enters $V_t$ only if the
action belongs to due-action set $D_t$ and its identifier is included in
acceptance set $A_t$. The quantity $|V_t|$ is the length of this list:
$|V_t|=0$ means that the transition produces no visible message, whereas
$|V_t|>0$ means that it produces at least one.

A deferred response must be distinguished from a decision not to respond. When
the system receives new user input but produces no visible message in the
current transition, it can create a $\mathsf{Reconsider}$ action associated with
that input and defer whether and how to respond. Formally, this corresponds to
$U_t\neq[]$, $|V_t|=0$, and a corresponding $\mathsf{Reconsider}$ in $N_t$.
The action does not promise that a message will eventually be produced. If the
system neither produces a visible message nor preserves such a future decision
opportunity, it declines to respond at the current decision point. Both plans
are valid, but they have different meanings.

For example, the system may refrain from replying immediately to a user message
and create a $\mathsf{Reconsider}$ action for ten minutes later. At that time it
decides from the latest context whether to reply or remain silent, rather than
writing a message in advance merely to display it later.

\subsection{Plan Revision and Due-Action Constraints}

Each future action $q$ has a stable unique identifier $\operatorname{id}(q)$.
For an action set $D_t$, $\operatorname{id}(D_t)=
\{\operatorname{id}(q)\mid q\in D_t\}$ denotes the set of identifiers of its
members. Let $\operatorname{targets}(\Delta_t)$ denote the identifiers targeted
by mutations. Because $A_t$ also contains action identifiers, a valid plan must
satisfy

\begin{align}
A_t &\subseteq \operatorname{id}(D_t), \\
A_t \cap \operatorname{targets}(\Delta_t) &= \varnothing, \\
\operatorname{id}(D_t) &\subseteq A_t \cup \operatorname{targets}(\Delta_t).
\end{align}

Here, $\varnothing$ denotes the empty set. The first constraint permits $A_t$ to
accept only currently due actions. The second requires $A_t$ and the revision
targets to be disjoint, preventing the same action from being both accepted and
revised by one plan. The third requires every due action to be either accepted
or explicitly revised. Revision targets must also be distinct and pending.
Consequently, each due action receives exactly one disposition in a transition:
an accepted $\mathsf{Send}$ enters $V_t$, an accepted $\mathsf{Reconsider}$ is
consumed, and every other action must be cancelled, replaced, or rescheduled.
These constraints prevent a due action from being omitted or handled in
conflicting ways.

Because $U_t$ and $D_t$ are observed together, new user input can invalidate a
due expression before delivery. Reaching an eligibility time permits
reconsideration; it does not independently authorize delivery.

For example, if the user reports that the interview was cancelled before or at
the instant when a short-delay suggestion becomes due, the new input and due
action must be handled by the same plan, and the obsolete suggestion must be
cancelled rather than delivered.

\subsection{Dialogue Pacemaking, Silence, and Renewed Contact}

When $w_t=1$, the planner receives another dialogue decision opportunity even if
there is no new user input or due queue action. This is the \emph{dialogue
pacemaker}. A valid plan with $|V_t|=0$ represents a deliberate choice to
remain silent rather than a planning failure. The pacemaker ensures that the
system can decide again, but it cannot prevent silence at every wakeup.

After a pacemaker decision completes, the control layer sets $P_{t^+}$ later
than $t$ using a configured positive interval, which may be fixed or adjusted
within prescribed bounds. A future wakeup may be retained or adjusted by a new
valid plan. Without user input, whichever comes first---a queued action becoming
due or the pacemaker wakeup---triggers the next decision. Thus, even an empty
queue leaves a defined opportunity to wake again.

To prevent repeated silence decisions from causing indefinite waiting in
proactive mode, we additionally define a bounded-silence policy. Let

\begin{equation}
L_t=\langle z_t,\ell_t,b_t\rangle,
\end{equation}

where $z_t$ indicates whether the user enables this policy, $\ell_t$ is the time
of the most recent visible AI message, and $b_t$ is the next contact deadline
set from user preferences and contact-rate rules. The symbol $\bot$ means that
no deadline currently exists; $b_t=\bot$ when the policy is disabled. On enabling
the policy, the control layer establishes the first deadline relative to the
activation time; subsequent visible messages reset this reference time. To express
whether the system has reached a decision point at which it may no longer remain
silent, define

\begin{equation}
g_t=z_t\land(b_t\neq\bot)\land(t\geq b_t)
\land\operatorname{allow}(t),
\end{equation}

where $\operatorname{allow}(t)$ denotes permission for autonomous contact at the
current time, supplied by an external policy that may account for user
authorization, quiet hours, rate limits, and domain safeguards. When there is no
pending user input ($U_t=[]$), visible output requires this permission; pausing
autonomous contact does not prevent the system from responding to user input.
Thus, $g_t$ is true only when the policy is enabled, a deadline exists and has been
reached, and external rules permit contact. Any plan that passes validation must
then satisfy $|V_t|>0$ rather than ending in silence. This constraint does not
predefine message content; the planner still determines what to say and how many
messages to produce from the then-current context. Earlier silence decisions
cannot clear $b_t$, and a failed plan does not discharge the contact requirement.
A new visible AI message updates $\ell_t$ and establishes the next deadline.
Temporary policy restrictions postpone contact; disabling the policy clears the
corresponding obligation.

The contact deadline also participates in wakeup scheduling. When contact is
permitted, a deadline earlier than the next regular pulse advances that pulse
to the deadline; this decision therefore does not depend on new input or an
existing queued action. If a temporary restriction is lifted after the deadline,
the system immediately requests a new decision. A failed plan preserves the
deadline and schedules another attempt under the recovery rules.

Before the deadline, a valid plan may remain silent. Once the deadline has
arrived and external rules permit contact, the next plan that passes validation
must produce visible output. Provided that the system continues running, external
rules permit contact, and planning eventually succeeds, this mechanism renews
contact rather than indefinitely postponing it through repeated acceptance of
silent plans.

\subsection{Plan Validation and Concurrent Arrivals}

A candidate plan is executed only if it satisfies the model constraints as a
whole. If it leaves a due action unresolved or proposes conflicting revisions,
the system neither delivers its messages nor applies its proposed additions,
cancellations, or revisions. Existing action contents, identifiers, and pending
statuses are preserved, and the associated user input remains unprocessed.
The control layer separately schedules a retry after a positive interval under
deterministic recovery rules, updating the next attempt times of actions and the
pacemaker accordingly. These recovery updates do not constitute partial execution
of the rejected plan or discharge an outstanding contact requirement.

User messages, due actions, and dialogue-pacemaker signals available before a
decision snapshot are first combined into event bundle $E_t$, from which the
system generates one plan. New context can therefore inform the decision before
pending content is delivered. Actions with the same timestamp are ordered by
stable identifiers.

A candidate plan is also bound to the state snapshot from which its context was
constructed; an implementation may enforce this condition with a monotonically
increasing state revision. If another user message arrives while the planner is
running, the current state no longer matches that snapshot. The older candidate
is rejected before atomic commit: it emits no visible message and makes no queue
mutation. The next decision includes both the earlier and newly arrived messages
in a new $U_t$. This rule lets independent sends within a short interval share
one decision while preventing inference latency from leaking a reply based on
incomplete context. Given the external events and the candidate that ultimately
passes validation, the control layer still produces a unique state transition.

\section{Research Prototype and Interaction Demonstration}

\subsection{System Implementation}

We implement a research prototype of Proactive AI that follows the temporal
semantics and state-transition constraints defined in Section~4. The prototype is open source,
and its code is available from the project repository.\footnote{%
\url{https://github.com/airalogy/proactive-ai}} The prototype separates
language planning from temporal execution. Planner $\pi$ is instantiated by a
language model, which proposes immediate messages, future actions, and queue
revisions as a structured candidate plan. Only candidates that pass structural
parsing and constraint validation may change the conversation state.

\begin{figure*}[t]
  \centering
  \resizebox{0.98\textwidth}{!}{\begin{tikzpicture}[pai-figure]
\tikzset{
  pai-arch-box/.style={
    rounded corners=3pt,
    draw=paiBorder,
    fill=white,
    line width=0.65pt,
    minimum height=0.92cm,
    align=center,
    inner xsep=5pt,
    inner ysep=4pt
  },
  pai-source-box/.style={
    pai-arch-box,
    draw=paiUser,
    fill=paiUserLight,
    text width=1.72cm,
    minimum height=2.25cm
  },
  pai-control-box/.style={
    pai-arch-box,
    text width=1.48cm
  },
  pai-planner-box/.style={
    pai-arch-box,
    draw=paiDecision,
    fill=paiDecisionLight,
    text width=2.18cm,
    minimum height=0.86cm
  },
  pai-commit-box/.style={
    pai-arch-box,
    draw=paiDecision,
    fill=paiDecisionLight,
    text width=1.50cm
  },
  pai-state-box/.style={
    pai-arch-box,
    draw=paiAgent,
    fill=paiAgentLight,
    text width=2.12cm,
    minimum height=0.95cm
  },
  pai-output-box/.style={
    pai-arch-box,
    draw=paiAgent,
    fill=paiAgentLight,
    text width=2.12cm,
    minimum height=0.72cm
  },
  pai-policy-box/.style={
    pai-arch-box,
    draw=paiNeutral,
    fill=paiPanel,
    text width=4.15cm,
    minimum height=0.78cm
  },
  pai-reject-box/.style={
    pai-arch-box,
    draw=paiRevision,
    fill=paiAgentLight,
    text=paiRevision,
    text width=2.75cm,
    minimum height=0.76cm
  },
  pai-flow/.style={
    -{Latex[length=1.8mm]},
    draw=paiNeutral,
    line width=0.68pt,
    rounded corners=2pt
  },
  pai-accepted/.style={
    -{Latex[length=1.8mm]},
    draw=paiDecision,
    line width=0.78pt,
    rounded corners=2pt
  },
  pai-rejected/.style={
    -{Latex[length=1.8mm]},
    draw=paiRevision,
    line width=0.78pt
  }
}

\filldraw[pai-panel] (0,0) rectangle (16.8,6.20);
\node[pai-panel-title, anchor=west] at (0.38,5.82)
  {Temporal plan validation and execution};

\node[pai-panel-title, anchor=west] at (0.29,4.95) {Event sources};
\node[pai-source-box] (sources) at (1.32,2.76) {
  \textbf{User arrivals $u_i$}\\[-1pt]
  {\scriptsize unprocessed $\rightarrow U_t$}\\[6pt]
  \textbf{Dialogue}\\[-1pt]
  \textbf{pacemaker $w_t$}\\[6pt]
  \textbf{Due}\\[-1pt]
  \textbf{actions $D_t$}\\[-1pt]
  {\scriptsize $0,\ldots,n$ actions}
};

\filldraw[rounded corners=5pt, fill=white, draw=paiBorder, line width=0.65pt]
  (2.45,1.48) rectangle (13.45,4.24);
\node[pai-panel-title, text=paiNeutral, anchor=west, align=left] at (2.58,3.88)
  {Deterministic\\[-2pt]control layer};

\node[pai-control-box] (dispatch) at (3.45,2.76)
  {1\\Form event\\bundle $E_t$};
\node[pai-control-box, text width=1.60cm] (snapshot) at (5.52,2.76)
  {2\\Take decision\\snapshot\\[-1pt]{\scriptsize $S_t,E_t$}};
\node[pai-control-box, text width=1.72cm] (validate) at (7.92,2.76)
  {3\\Validate\\plan\\[1pt]{\scriptsize structure + rules\\[-3pt]+ contact policy}};
\node[pai-control-box, text width=1.68cm] (draft) at (10.17,2.76)
  {4\\Prepare next\\state\\[-1pt]{\scriptsize temporary copy}};
\node[pai-commit-box] (commit) at (12.42,2.76)
  {5\\Finalize\\transition\\[-1pt]{\scriptsize all or none}};

\node[pai-planner-box] (planner) at (5.52,5.05)
  {Planner $\pi$\\[-1pt]{\scriptsize LLM-based}};
\node[pai-policy-box] (policy) at (10.65,5.05)
  {\mbox{Autonomous-contact policy}\\[-1pt]
   {\scriptsize $\operatorname{allow}(t)$}};

\node[pai-panel-title, anchor=east] at (16.36,4.95) {Transition results};
\node[pai-output-box] (output) at (15.48,3.43) {
  \textbf{Visible output}\\[-1pt]
  {\scriptsize $V_t$: $0,\ldots,n$ messages}
};
\node[pai-state-box] (state) at (15.48,1.95) {
  \textbf{Updated state}\\[1pt]
  {\scriptsize $S_{t^+}$}
};

\node[pai-reject-box] (reject) at (7.92,1.02)
  {Candidate rejected\\[1pt]{\scriptsize no output;\\[-3pt]candidate not applied}};

\draw[pai-flow] (sources.east) -- (dispatch.west);
\draw[pai-flow] (dispatch.east) -- (snapshot.west);
\draw[pai-accepted] (validate.east) -- (draft.west);
\draw[pai-accepted] (draft.east) -- (commit.west);
\coordinate (resultfork) at (13.82,2.76);
\draw[pai-accepted] (commit.east) -- (resultfork);
\draw[pai-accepted] (resultfork) |- (output.west);
\draw[pai-accepted] (resultfork) |- (state.west);

\draw[pai-flow]
  (snapshot.north) -- (planner.south);
\draw[pai-flow]
  (planner.east) -- (7.92,5.05) -- (validate.north);
\node[pai-secondary, align=center] at (7.92,5.59)
  {candidate temporal\\plan};
\draw[pai-flow]
  (policy.south) -- ++(0,-0.20) -| ([xshift=7pt]validate.north);

\draw[pai-rejected] (validate.south) -- (reject.north);
\draw[pai-flow, densely dashed]
  (state.south) -- (15.48,0.34) -- (5.52,0.34) -- (snapshot.south);
\node[pai-secondary, fill=paiPanel, inner xsep=3pt] at (11.25,0.34)
  {state for the next event};

\end{tikzpicture}}
  \caption{Architecture for temporal plan validation and execution. The diagram
  reads from left to right as one state transition. Each $u_i$ records one
  independent user send. At the start of a decision, messages not yet handled by
  a successfully committed plan form the ordered list $U_t$. This list is combined with dialogue-
  pacemaker signal $w_t$ and due actions $D_t$, namely future actions whose
  execution time has arrived, to form event bundle $E_t$. The LLM-based planner
  $\pi$ reads the current state $S_t$ and $E_t$, then proposes a candidate
  temporal plan. Such a plan may contain immediate messages, new future actions,
  revisions to the existing queue, and decisions about due actions. Once the
  candidate plan has been produced, the control layer validates it and updates
  the state according to fixed rules. It checks that the plan is structurally
  complete, still refers to the current state, and preserves state and queue
  consistency. For visible output without pending user input, it also checks
  autonomous-contact permission $\operatorname{allow}(t)$. A valid plan is
  first applied to a temporary state.
  Only after every check succeeds do visible output $V_t$ and next state
  $S_{t^+}$ take effect together; otherwise neither takes effect. $V_t$ may
  contain zero, one, or many messages. $S_{t^+}$ retains the updated history,
  future-action queue, pacemaker state, and contact policy for the next event,
  as shown by the dashed path. If a candidate plan does not satisfy these
  execution conditions, its messages and proposed state changes do not take
  effect. Unprocessed input and due actions are retained; the control layer may
  separately update retry times for a later decision based on the latest state.}
  \Description{Independent user-message arrivals, a dialogue-pacemaker signal,
  and due actions enter a five-stage deterministic control layer. An LLM-based
  planner above the layer receives the current-state decision snapshot and
  returns a candidate temporal plan; an autonomous-contact-policy box also feeds the
  validation stage. The successful path validates the plan, prepares a temporary
  next state, then finalizes visible output and the updated state together. A
  dashed arrow carries that state to the next decision. A red branch marks a
  candidate whose output and proposed changes are not applied; subsequent retries
  follow separate recovery rules.}
  \label{fig:architecture}
\end{figure*}
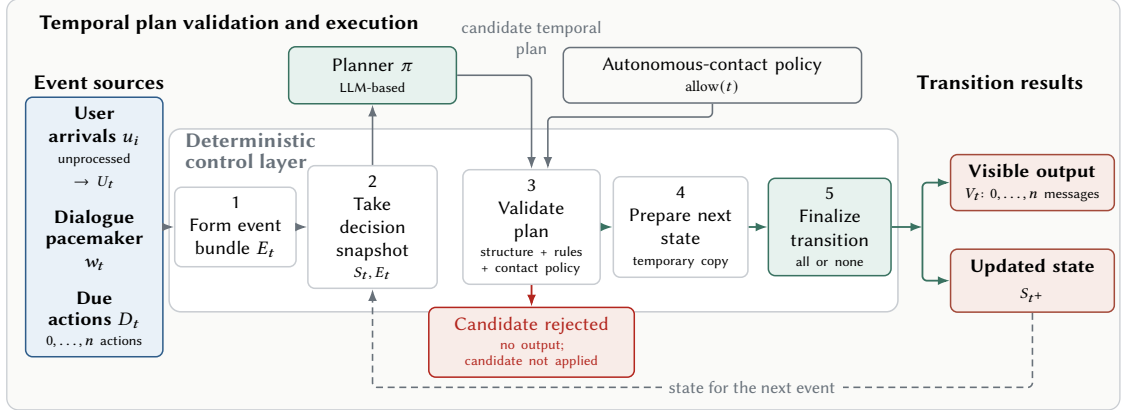

An input coordinator records every user send as an independent arrival, groups
nearby unprocessed messages into a decision snapshot, and discards an older
candidate if the state changes during inference. A deterministic control layer
owns wall-clock time, validates candidate plans, updates the future-action queue,
maintains the bounded-silence deadline, enforces the externally supplied
permission for autonomous contact, and finalizes visible output and state
changes. The interface aligns the visible
conversation with pending future actions and the event trace that produced them,
making the temporal state available for inspection.
Figure~\ref{fig:architecture} presents the resulting architecture.

\subsection{Illustrative Interaction}

\begin{figure*}[t]
  \centering
  \includegraphics[width=\textwidth]{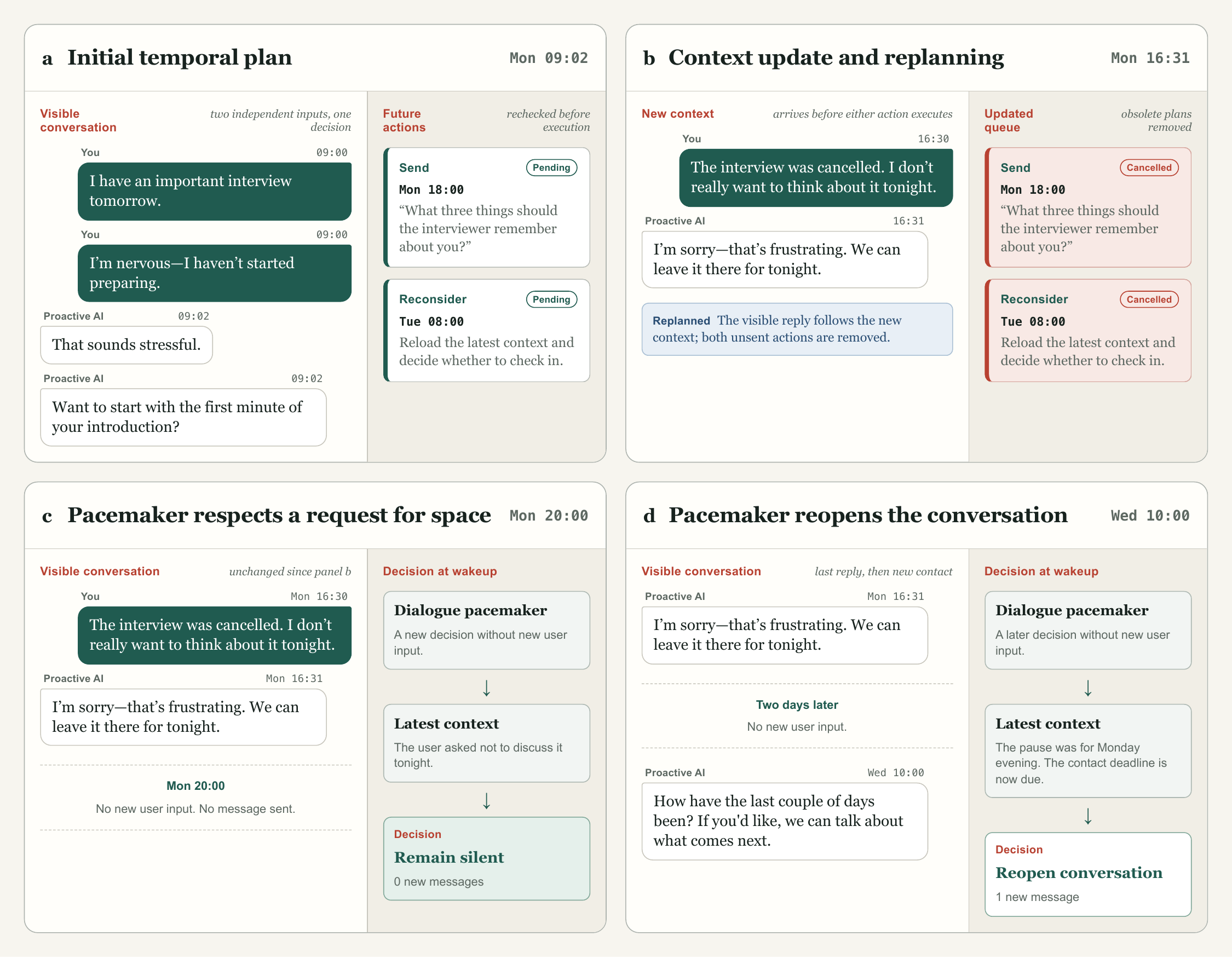}
  \caption{An illustrative temporal dialogue.
  \textbf{a},~Two independently sent user messages are observed in one decision, which
  yields two immediate messages and two types of future action: \emph{Send}
  stores the quoted message text, whereas \emph{Reconsider} retains only the
  purpose of a later decision about whether to follow up.
  \textbf{b},~After the user reports that the interview was cancelled, both unsent
  actions are cancelled internally, while the visible reply responds to the new
  situation rather than describing queue operations. \textbf{c},~A pacemaker trigger
  later that evening respects the user's request for space: the earlier exchange
  remains visible, but no new message is added. \textbf{d},~Two days later, another
  pacemaker trigger uses the latest context to initiate a check-in without new
  user input, shown as a new chat message after the time gap. The chat history
  and the corresponding decision are shown side by side. Together, the final two
  states show that a decision opportunity can yield either silence or renewed
  contact.}
  \Description{Four states of an illustrative temporal dialogue. The first state
  shows two immediate assistant messages and two pending future actions. The
  second shows a later user update, a natural acknowledgement, and both future
  actions cancelled internally. The third pairs the unchanged earlier chat with
  an evening pacemaker decision that remains silent because the user requested
  space. The fourth shows the last earlier reply, a two-day gap without user
  input, and a new check-in message beside the corresponding pacemaker decision.}
  \label{fig:prototype}
\end{figure*}

Figure~\ref{fig:prototype} uses an interview dialogue to show
user-visible messages alongside the corresponding future actions and decision
states. In the first state, two
independently sent user messages are observed in one decision, which produces a
two-message assistant response, a prepared message for later delivery, and a later
reconsideration. After the user reports that the interview was cancelled, the
next decision removes both future intentions before either can be delivered and
responds naturally to the updated situation. A dialogue-pacemaker event later
that evening respects the user's request for space and remains silent. Two days
later, another pacemaker event reopens the conversation with a check-in despite
the absence of new user input.

\section{Discussion}

The proposed model allows dialogue to continue without depending entirely on
another user request. Even without a pending message or a scheduled follow-up,
dialogue pacemaking lets the system use the available context to decide whether
to initiate an exchange. It can return to unfinished matters or raise a new
topic after an earlier exchange has ended. The system can therefore share the
work of remembering unresolved matters, choosing follow-up times, and initiating
further exchanges, creating an opportunity to reduce the user's coordination
burden in sustained collaboration. Continuity need not take the form of
continuous messaging; it can also consist in timely participation after a period
of silence.

From this perspective, systems for longitudinal collaboration need not only to
retain information about the past but also to organize their future participation.
Memory helps the system understand what has already happened; temporal plans
represent what merits later attention and when another decision should be made.
The two are complementary: past context informs future actions, while those
actions must be revised as new information arrives. The system can thus retain
an intention to follow up without committing in advance to what it will say.

For example, an AI tutor could revisit an unresolved difficulty after an
explanation and use the learner's subsequent progress to decide whether to
explain further, adjust an exercise, or refrain from intervening. A scientific
assistant could retain a question for reconsideration while awaiting experimental
results, then decide whether to continue the analysis or revise a suggestion when
new evidence becomes available. In these settings, the timing and form of the
system's participation should adapt to task progress.

Beyond task-oriented collaboration, relational-agent research treats the
establishment and maintenance of human--computer relationships across repeated
interactions as a design concern \cite{bickmore2005relationships}. The proposed
temporal organization could support AI companions and other systems for everyday
conversation and emotional support. In these settings, humanlike
interaction involves not only language style but also initiative, continuity,
and sensitivity to changing circumstances. Without another user question, the
system can draw on prior exchanges to initiate a supportive check-in, adapting
both its timing and content to the latest context. For example, after learning
that the user's interview was cancelled, it could respect a request not to
discuss it for the moment, withdraw the original interview reminders, and ask
how the user is doing at a suitable later time. Considerate interaction in this
sense involves more than warm wording: it requires attention to the user's
current circumstances when choosing between initiating contact and allowing
space.

Proactive dialogue also offers a starting point for studying autonomous behavior
more broadly. In their work on AI scientists, Yang et al. distinguish the
initiation of scientific inquiry from the execution of research tasks and
discuss whether AI could assume the role of the ``First Mover of Science''
\cite{yang2023ai}. This distinction suggests that autonomously completing a
given task and autonomously initiating an activity are different dimensions of
capability. Starting from sustained dialogue, the present work brings renewed
decisions without a new user request and context-dependent revisions of future
actions into an executable temporal model. This mechanism can serve as a basis
for research into broader forms of autonomous interaction, with potential
extensions to high-level task planning for collaborative robots.

For example, within an authorized laboratory inspection task, a robot could
independently schedule a later visit to a device to check its status and
reassess the need for that visit before execution. If a staff member has already
completed the check, it could cancel the unexecuted visit; if the passage is
temporarily occupied, it could postpone the action. Such an extension would
require integrating temporal plans with environmental perception, action
execution, and safety controls, keeping a renewed decision opportunity distinct
from the execution of a physical action.

In summary, the proposed model brings decisions about whether to participate,
when to participate, and how to revise participation plans as context changes
into one conversational process. Its significance extends beyond a more natural
messaging rhythm to a temporal organization of sustained collaboration in which
participation can be initiated, deferred, and revised. The value of initiative
should lie in timely support for user needs and shared activities, rather than
in the number of messages or actions produced.

\section{Conclusion}

This paper introduces Proactive AI, a framework for proactive dialogue built
around replannable temporal message queues. It models conversation as a
continuous event timeline and represents silence, message bursts, delayed
replies, and proactive initiation within a single model.
By representing unexecuted conversational actions as revisable state and
reconsidering them against the latest context before execution, the model allows
future expressions to be revised or withdrawn rather than automatically
delivered when their scheduled time arrives. The formal model and open-source
research prototype provide an executable basis for this interaction paradigm.
Its significance extends beyond the rhythm of message delivery: AI can reinitiate
exchanges without a new request, maintain attention to unfinished matters, and
adapt subsequent actions to changing circumstances. Conversational AI can thus
move from individual responses toward sustained, revisable participation,
providing a mechanism for sharing the work of follow-up and coordination in
long-term human--AI collaboration. This organization of future participation also
offers a starting point for studying broader forms of autonomous interaction.

\begin{acks}
The author thanks Shufeng Sun, a roommate during the author's doctoral studies,
for helpful discussions on interaction scenarios for proactive conversational AI.
\end{acks}

\bibliographystyle{ACM-Reference-Format}
\bibliography{references}

\clearpage
\appendix
\section{Supplementary Figures}
\label{app:supplementary-figures}
\setcounter{figure}{0}
\renewcommand{\theHfigure}{supp.\arabic{figure}}
\captionsetup[figure]{name=Supplementary Figure}

\begin{figure*}[ht]
  \centering
  \includegraphics[width=\textwidth,height=0.78\textheight,keepaspectratio]{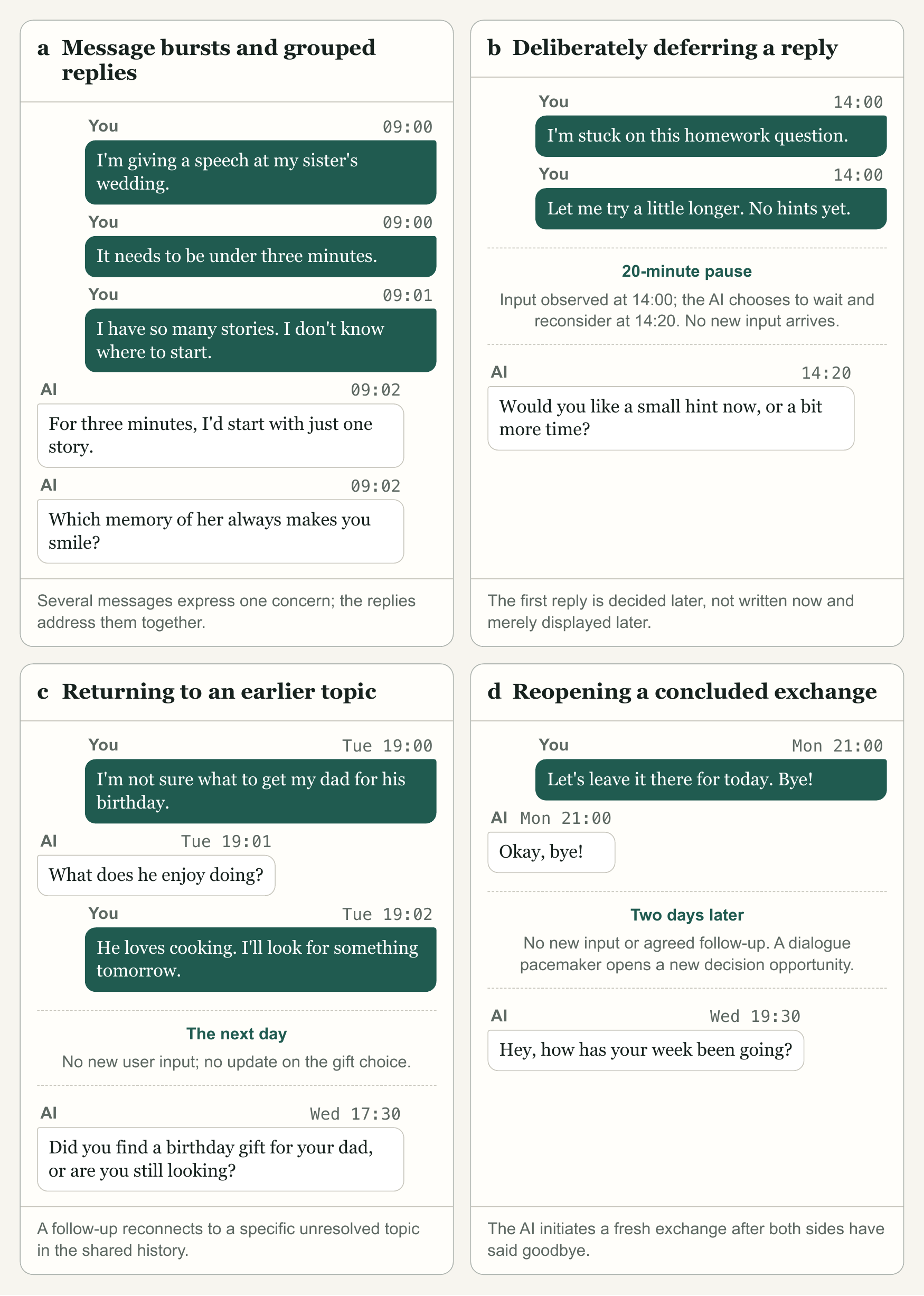}
  \caption{Illustrative dialogue patterns beyond paired turns.
  \textbf{a},~Consecutive inputs and a combined response. Several user messages
  jointly express one concern, addressed by successive AI replies rather than
  one answer per input.
  \textbf{b},~A delayed first reply. The AI observes the user input at 14:00 but
  initially remains silent, waiting until 14:20 to reconsider and generate its
  first reply.
  \textbf{c},~Following up across time. The AI later returns to a previously
  discussed but unresolved topic, continuing the earlier exchange.
  \textbf{d},~Proactive initiation after goodbye. After both sides say goodbye,
  the dialogue pacemaker gives the AI a new decision opportunity, leading it to
  initiate a new exchange without new user input or an agreed follow-up.}
  \Description{Four chat panels with timestamps. Panel a shows three user
  messages about a short wedding speech followed by two AI replies.
  Panel b shows a user requesting time to solve a homework question, an annotated
  twenty-minute deliberate pause, and the first AI reply.
  Panel c shows an exchange about choosing a birthday gift for the user's father
  and an AI follow-up the next day. Panel d shows both sides saying goodbye and the AI starting a
  new exchange two days later. User messages are right aligned and dark green;
  AI messages are left aligned and white. Decision annotations are outside
  message bubbles.}
  \label{fig:dialogue-patterns}
\end{figure*}

\end{document}